 \documentclass[review,5p]{elsarticle}

\usepackage{amssymb}
\usepackage{amsmath, nccmath}
\usepackage{lipsum}

\usepackage[normalem]{ulem}
\usepackage{color}
\usepackage{float}
	
\usepackage{microtype}
\usepackage{hyperref}
\renewcommand{\epsilon}{\varepsilon}

\journal{NIM-A}
\begin{document}

\begin{frontmatter}



\title{Observation of light production by charged particles in WLS fibers}



\author[a,b,e]{I.~Alekseev,}

\author[b]{A.~Chvirova,}

\author[a]{M.~Danilov}

\author[b]{S.~Fedotov,}

\author[b]{A.~Khotjantsev,}
\author[b]{M.~Kolupanova,}
\author[d]{N.~Kozlenko,}
\author[a]{A.~Krapiva,}
\author[b,c,f]{Y.~Kudenko,}

\author[a,b]{A.~Mefodiev,}
\author[b]{O.~Mineev,}

\author[d]{D.~Novinsky,}

\author[a,e]{V.~Rusinov}

\author[a,e]{E.~Samigullin,}
\author[a,b,e]{N.~Skrobova,\corref{correspondingauthor}}
\cortext[correspondingauthor]{Corresponding author: skrobovana@lebedev.ru}
\author[a,b,e]{D. Svirida,}
\author[e]{E.~Tarkovsky.}


\affiliation[a]{organization={Lebedev Physical Institute of the Russian Academy of Sciences}, addressline={Leninskiy avenue 53, Moscow, 119991, Russia}}
\affiliation[b]{organization={Institute for Nuclear Research of the Russian Academy of Sciences}, addressline={60th October Anniversary Prospect 7a, Moscow 117312, Russia}}
\affiliation[c]{Moscow Institute of Physics and Technology, addressline={Institutskiy lane 9, Dolgoprudny, Moscow Region, 141701, Russia}}
\affiliation[d]{organization={B.P. Konstantinov Petersburg Nuclear Physics Institute}, addressline={mkr. Orlova rocha 1, Gatchina, Leningrad Oblast, 188300, Russia}}
\affiliation[e]{organization={National Research Center ``Kurchatov Institute''}, addressline={Akademik Kurchatov square 1, Moscow, 123182, Russia}}
\affiliation[f]{organization={National Research Nuclear University Moscow Engineering Physics Institute}, addressline={Kashirskoe shosse 31, Moscow, 115409, Russia}}

\begin{abstract}
Wavelength shifting (WLS) fibers are widely used in particle physics for light collection from scintillators. Light production by charged particles directly in WLS fibers is traditionally ignored. In this study, light produced by charged particles in WLS fibers is clearly observed. 
The light yield of different batches of Y11(200) 1 mm diameter WLS fibers expressed in the number of detected photo electrons is as large as $23\pm2~\%$ with respect to the light yield of the Bicron BCF-12 1 mm diameter scintillating fiber. This corresponds to about 1200 photons per MeV light production after taking into account different fiber light trapping efficiencies and emission spectra.
In clear fibers of the same diameter, no scintillation light is produced, while Cherenkov light is clearly seen at the 45-degree crossing angle.
The observed amount of light produced by charged particles in the WLS fibers is not small and should be taken into account in advanced detector simulations. 
\end{abstract}


\end{frontmatter}




\section{Introduction}

Wavelength shifting (WLS) fibers are widely used in particle physics and other applications for light readout from scintillators. For example, about 8~km of WLS fibers were used in the DANSS detector~\cite{DANSS:design} and more than 70~km of WLS fibers were used in the SFGD detector~\cite{SFGD:design, T2K:2026zms, Kudenko:2025dlg}.  
Traditionally, light produced directly by charged particles in WLS fibers is not considered in simulations. It is assumed to be negligible (see e.g. simulations of the JUNO TAO muon system~\cite{TAO-Muon2302.12669v2}). However, this assumption is not completely correct. Sizable signals were observed from WLS fibers not connected to scintillators in two detector prototype tests at a 730~MeV/c pion beam at the SC-1000 synchrocyclotron (PNPI, Gatchina, Russia). 

\begin{figure}[!htb]
\centering
\includegraphics[width=0.5\textwidth]{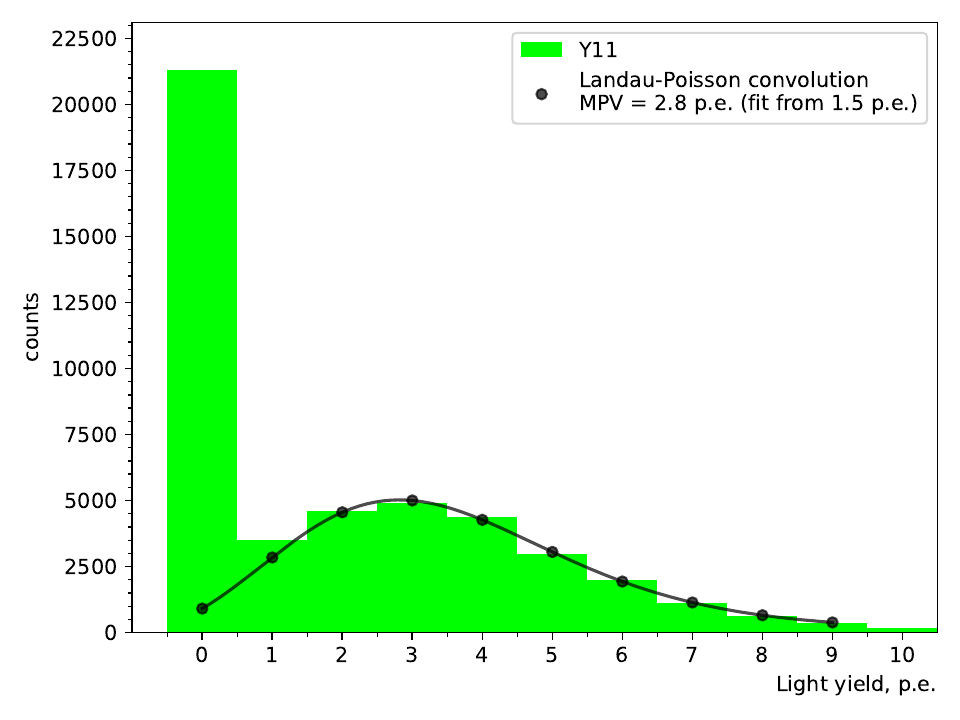}
\caption{\label{fig:WLS_LY} Light yield from 1.2~mm diameter Y11(200)M WLS fiber for tracks with a calculated track length in the fiber above 1.0~mm. The black curve shows the convolution of the Landau distribution for different track lengths and the discrete Poisson statistics.}
\end{figure}

One prototype was a net of 4$\times$4 1.2~mm diameter Kuraray Y11(200)M WLS fibers perpendicular to the beam, running along the Z coordinate with a distance of 15 mm between the fibers in the X and Y coordinates. Fibers were read out with Silicon Photo-multipliers (SiPM) (Hamamatsu MPPC S13360-1350PE). Fiber open ends were painted with  black paint. Signals from MPPCs were read out with custom-made digitizers developed for the DANSS experiment~\cite{DANSS:electronics}.  Initially,  the space between fibers was empty (just air). Still, clear signals from the WLS fibers were observed when the beam crossed them. The beam position was determined with an accuracy of about 0.5~mm  using a set of proportional chambers. Details of the beam setup are presented elsewhere~\cite{aquarium:beam}. When the calculated track length in the fiber was above 1.0~mm, the most probable value of the light yield (LY) was about 3 photo-electrons (p.e.), as shown in Fig.~\ref{fig:WLS_LY}. The actual track length had a spread because the accuracy of the track position determination was only about twice as good as the fiber diameter.   When the volume between fibers was filled with about 3~mm diameter plastic scintillator granules (polystyrene doped with 2\% P-terphenyl (PTP) and 0.02\% POPOP
(1,4-bis(5-phenyl-2oxazolil)benzene) fluors) with a density of 0.67~g/cm$^3$, the most probable light yield for tracks crossing the fibers increased to about 39 p.e. This demonstrates that the direct signals from the WLS fibers are not negligible.

\begin{figure}[!htb]
 \centering
\includegraphics[width=0.5\textwidth]{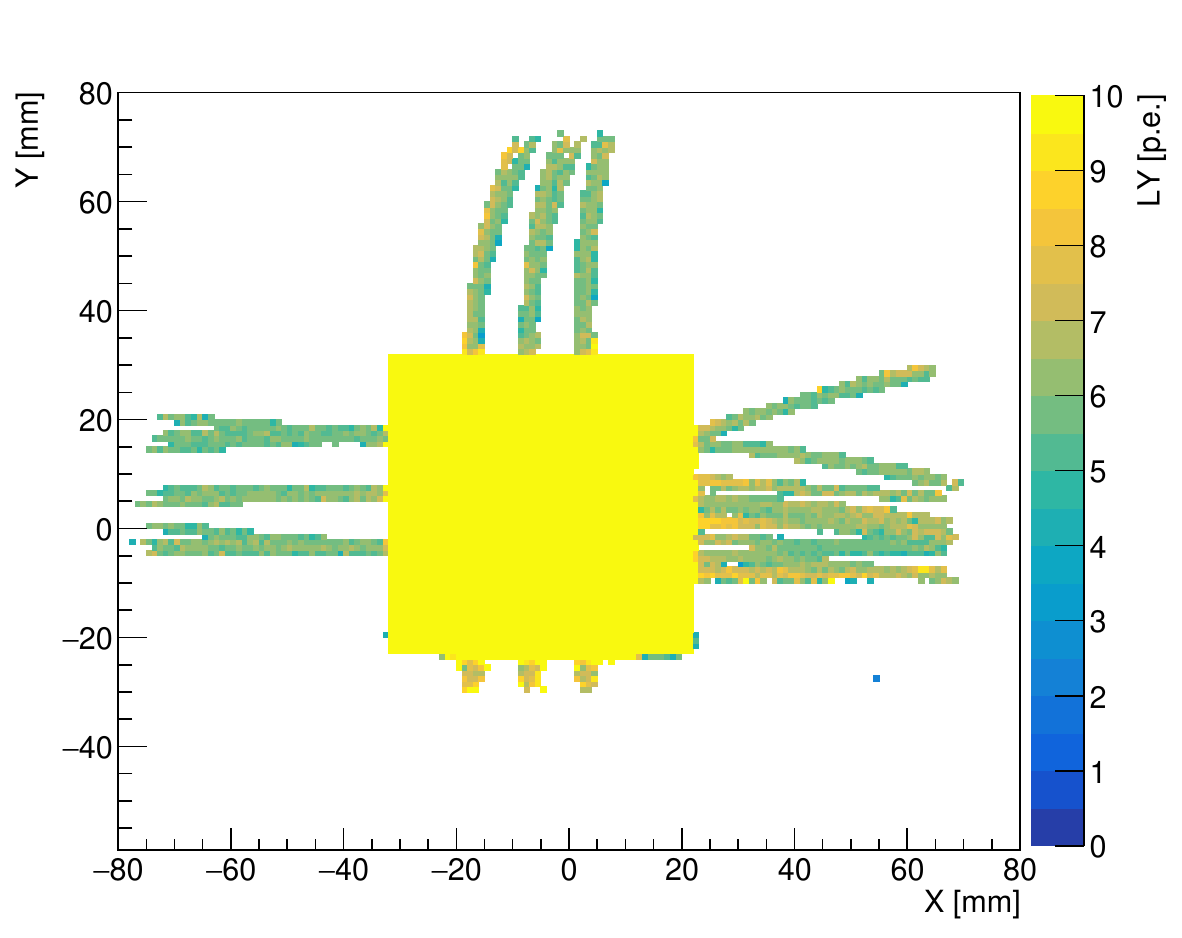}
\caption{\label{fig:HitMap} Light yield in p.e. obtained with the SFGD 5$\times$5$\times$5 scintillator cube prototype. Yellow region includes also entries with LY above 10 p.e. WLS fibers outside the scintillator volume are clearly seen.  }
\end{figure}

\begin{figure}[!htb]
\centering
\includegraphics[width=0.5\textwidth]{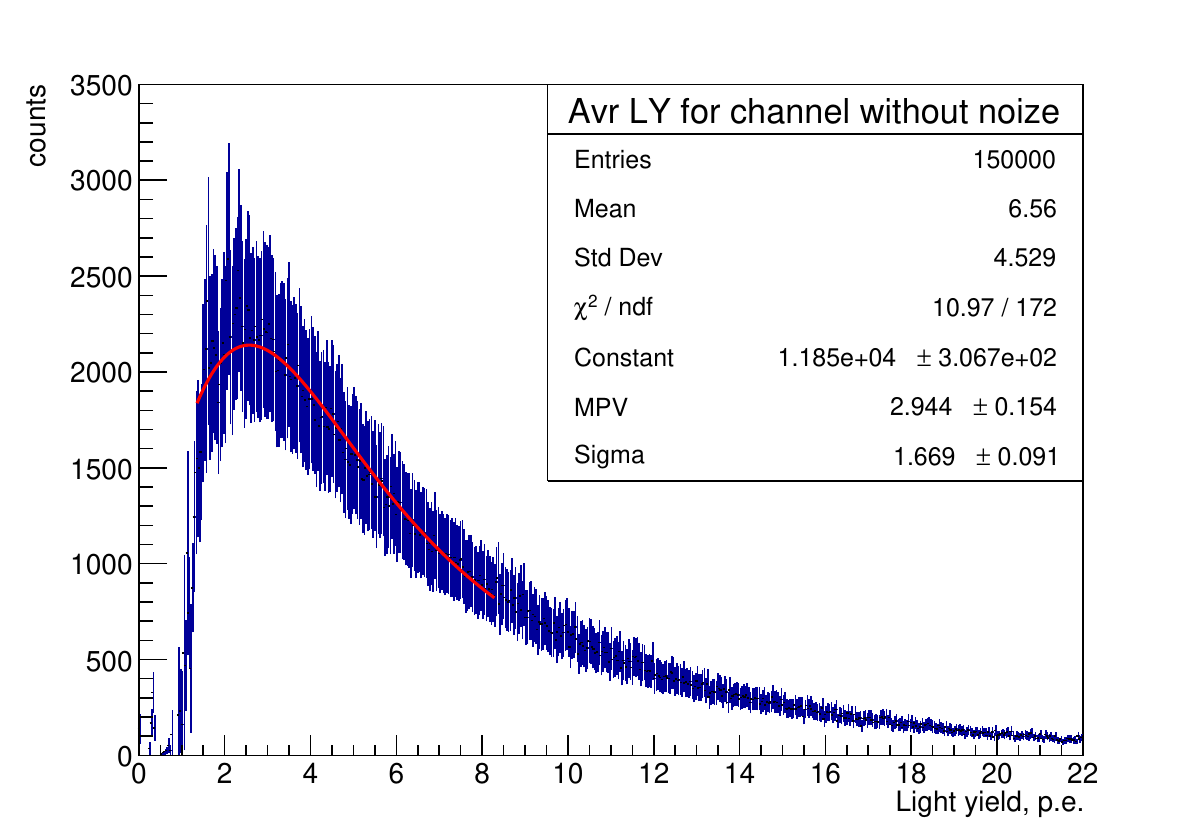}
\caption{\label{fig:WLS_LY_SFGD} Light yield produced directly by 730~MeV/c pions in all instrumented  1~mm diameter Y11(200)MSJ fibers of the SFGD prototype. Curve shows results of the fit with the Landau distribution. Since the light yield for particles traversing the WLS fiber is only about 3 p.e., the observed distribution is significantly influenced by photoelectron counting statistics. Therefore, the most probable value (MPV) obtained from the Landau fit should be interpreted as an effective estimator of the peak position rather than the most probable value of an ideal continuous Landau distribution.
}
\end{figure}

Another prototype was a 5$\times$5$\times$5 scintillator cube SFGD prototype~\cite{alekseev2026precisionlightyieldcrosstalk}. Each cube had a size of  one cm$^3$ and 3 orthogonal 1.5~mm  diameter holes for fiber insertion. Central 3$\times$3$\times$3 cubes of this prototype were read out in X,Y, and Z coordinates using 27 Kuraray Y11(200)MSJ 1~mm diameter WLS fibers connected to SiPMs (Hamamatsu MPPC S12571-025C). SiPMs were read out with the DANSS digitizers~\cite{DANSS:electronics}.
The WLS fibers outside the 5$\times$5$\times$5 scintillator cube array are clearly seen in the light yield map obtained after noise subtraction (see Fig.~\ref{fig:HitMap} ). The signal amplitude distribution after the SiPM noise subtraction is shown in Fig.~\ref{fig:WLS_LY_SFGD}. The most probable value was $(2.9\pm0.15)$~p.e. (statistical errors only) while the average value was $6.9$~p.e.
These numbers can be compared with the most probable LY of about 50~p.e. per fiber produced by a minimum ionizing particle (MIP) crossing the scintillator cube\cite{alekseev2026precisionlightyieldcrosstalk}. Obviously, one should take into account that only 20\% of tracks crossing the cube also cross the fiber. Still, the direct WLS fiber contribution to LY is not completely negligible. 

In order to obtain more quantitative estimates of the light produced by charged particles directly in WLS fibers, measurements with a radioactive source were performed. They are presented in the next section.    

\section{Measurements of LY produced in WLS fibers by electrons from a radioactive source}

\begin{figure}[!htb]
\centering
\includegraphics[width=0.5\textwidth]{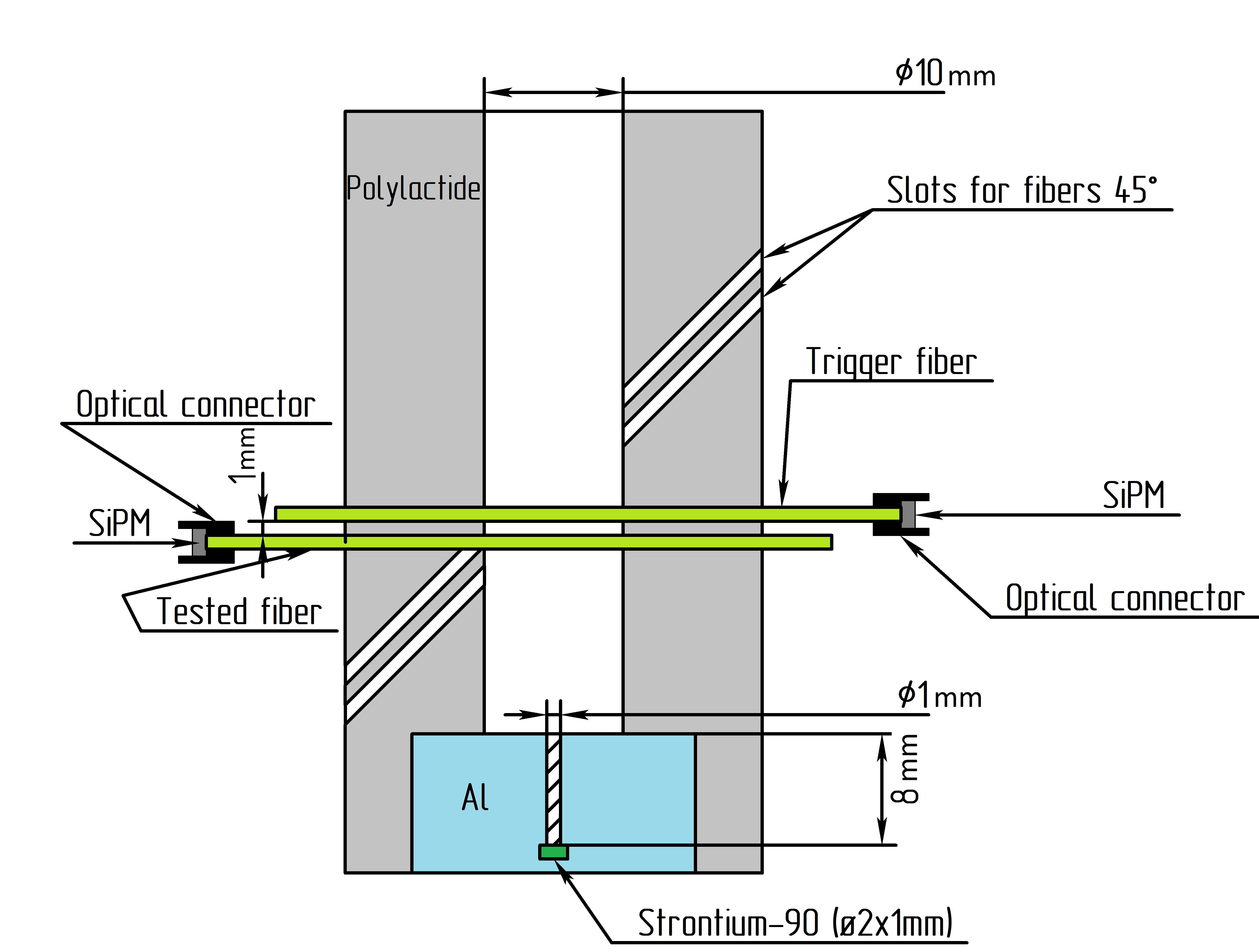}
\caption{\label{fig:Setup} Setup for fiber LY measurements using a $^{90}Sr$ radioactive source (not to scale).}
\end{figure}

The LY produced in WLS fibers by electrons from a $^{90}$Sr radioactive source was measured using a trigger from a scintillating fiber. A sketch of the setup is shown in Fig.~\ref{fig:Setup}. A 5~mCi $^{90}$Sr radioactive source was placed in an 8~mm long aluminum collimator with a 1~mm diameter hole. The collimator was placed in a 3D printed polyactide support structure with a 10~mm diameter hole and slots for two fibers perpendicular to the collimator and two slots at 45$^{\circ}$.  The slots had a diameter of 1~mm. The distance between the radioactive source and a tested fiber was 22~mm for the orthogonal fiber position. The trigger scintillating fiber (a single clad 1~mm diameter BCF-12) was placed behind a tested fiber. The trigger fiber was painted with black paint to prevent cross-talk with the studied fiber. The trigger and tested fibers were read out with SiPMs (Hamamatsu MPPC S13360-1350PE at 4~V overvoltage) and digitized using a 16-channel 12-bit ADC~\cite{IHEP_ADC} with a sensitivity of 0.25~pC/bin. One p.e. corresponds to about 15 ADC counts.
All tested fibers were read out with the same SiPM. To prevent damage to the SiPM surface from multiple insertions of fibers, a 20~$\mu$ thick polyethylene film was placed in front of the SiPM. The fibers and polyethylene film were air coupled.
The film reduced LY by about 10\%. During insertion, the fibers were pressed to touch this film.
The fibers had a  length of about 20~cm. The trigger threshold was 12 p.e. All tested fibers had a 1~mm diameter. The measurements were performed at a room temperature of about 22~$^{\circ}$C. 
The SiPM cross-talk was measured to be about (9--10)\% using spectra produced with a light emitting diode. The method of the cross talk determination was described in~\cite{Balagura:2005gh}.
The reproducibility of the measurements was studied by performing multiple (2--14) measurements  with the same fiber at different times, including the fiber insertions into the optical connector. The standard deviation of the results was taken as a rough estimate of the systematic uncertainty in the fiber average LY.  It varied between 7\% and 16\% for WLS and scintillating fibers, as can be seen in Table~\ref{tab:all_angles}.

\begin{figure}[!htb]
\centering
\includegraphics[width=0.5\textwidth]{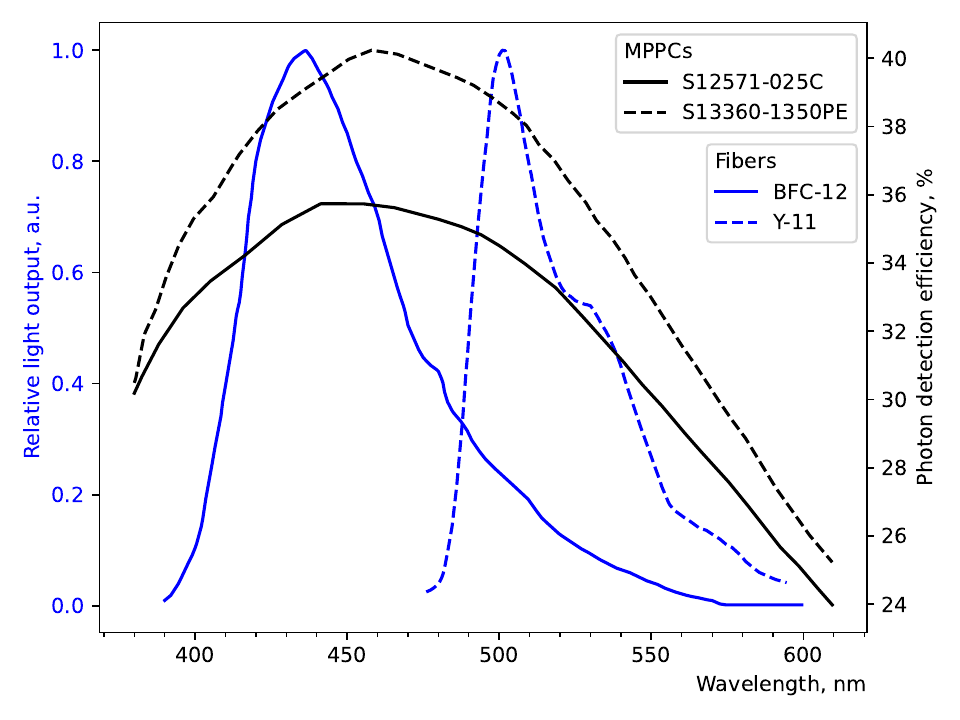}
\caption{\label{fig:spectra} 
Emission spectra of Y11 (dashed blue curve) and BCF-12 (solid blue curve) fibers compared with SiPM efficiencies at the nominal overvoltage.
}
\end{figure}

The following fibers were studied. A double clad Kuraray WLS fiber Y11(200)MSJ used in the Baby-MIND detector~\cite{Baby-MIND} (hereafter referred to as ``Y11-1'' fiber), two samples of the  double clad Kuraray WLS fiber Y11(200)MS used in the SFGD detector~\cite{SFGD:design} (hereafter referred to as ``Y11-2'' and ``Y11-3'' fibers), a single clad scintillating Bicron fiber BCF-12, and a clear Bicron fiber BCF-98.
The emission spectra of the Y11 and BCF-12 fibers are shown in Fig.~\ref{fig:spectra} together with the SiPM efficiencies.

\begin{figure}[!htb]
\centering
\includegraphics[width=0.5\textwidth]{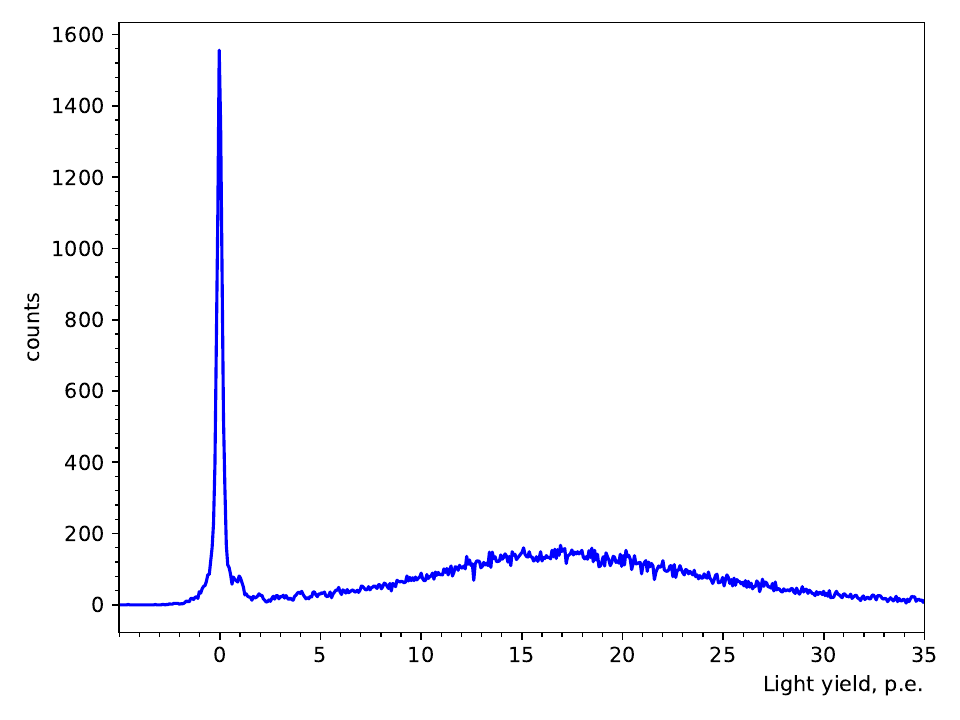}
\caption{\label{fig:Scintillating}  Light yield of BCF-12 scintillating fiber from a triggered $^{90}$Sr radioactive source for a 90$^{\circ}$ crossing angle.}
\end{figure}

Signals from the scintillating fiber (BCF-12) for the orthogonal fiber position are quite large, with an average value of about 18~p.e. (see Fig.~\ref{fig:Scintillating}). The number of p.e. was calculated from  the number of fired SiPM pixels, corrected for the cross-talk.  

\begin{figure}[!htb]
\centering
\includegraphics[width=0.5\textwidth]{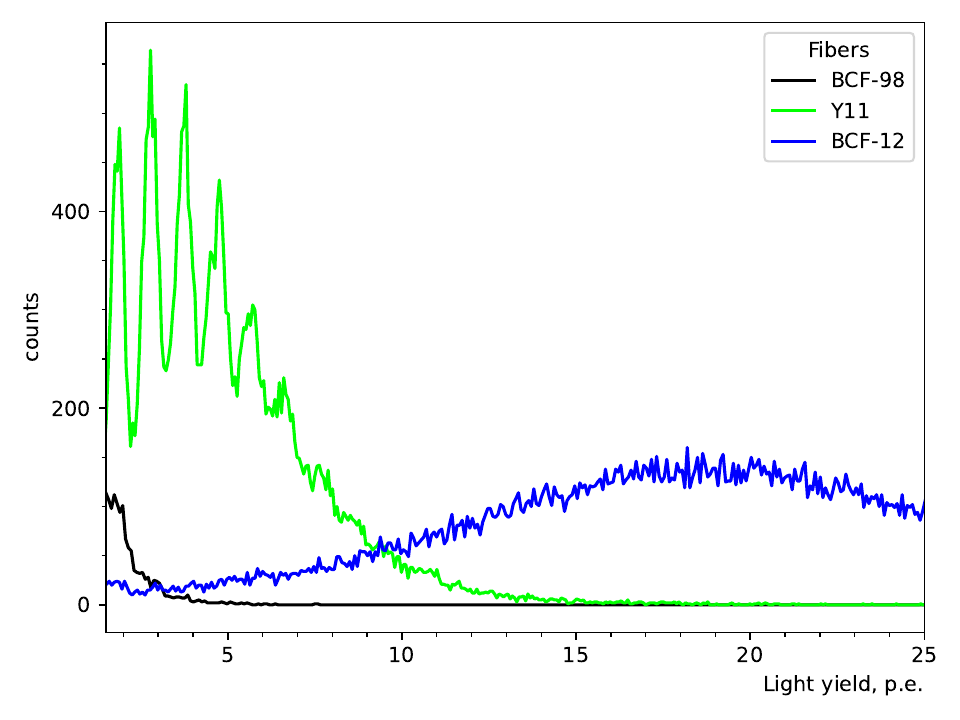}
\caption{\label{fig:90degree}  Light yield in different fibers from a triggered $^{90}Sr$ radioactive source for a 90$^{\circ}$ crossing angle. The black, green, and blue histograms correspond to the clear, WLS, and scintillating fibers. The numbers of entries for the clear and WLS fibers in the range from -2~p.e. to 200~p.e. were normalized to the  number of entries for the scintillating fiber  and the pedestal peak observed for the scintillating fiber was subtracted. The distributions are shown above 1.5~p.e.}
\end{figure}

Signals in the pedestal peak ($\pm0.5$~pixels)
are predominantly due to tracks that cross the trigger fiber but do not cross the tested fiber. Therefore, they are discarded in the LY calculations. Moreover, it is assumed that the fraction of such tracks is the same in the measurements of the WLS and clear fibers. Therefore,  the signal distribution between $-$0.5 and $+$0.5 pixels obtained in the measurements with the scintillating fiber was subtracted from the signal distributions obtained for the WLS and clear fibers after normalization to the total number of triggers for the scintillating fiber. 
The obtained LY distributions for the orthogonal position of the fibers are shown in Fig.~\ref{fig:90degree}. The distributions for the  ``Y11-1'',  ``Y11-2'', and ``Y11-3'' fibers are very similar. Therefore, results for the ``Y11-3'' fiber only are presented in the  figures below. The LY distributions are shown for one particular set of measurements. As explained above, several sets of measurements (typically 5) were performed to estimate the reproducibility of the results. 

There is practically no signal from the clear fiber. The dominant part of the distribution is in the pedestal peak below 0.5~p.e. The average LY is about 0.1~p.e. only. The three WLS fibers give an average signal of $3.81\pm0.24$~p.e.

\begin{figure}[!htb]
\includegraphics[width=0.5\textwidth]{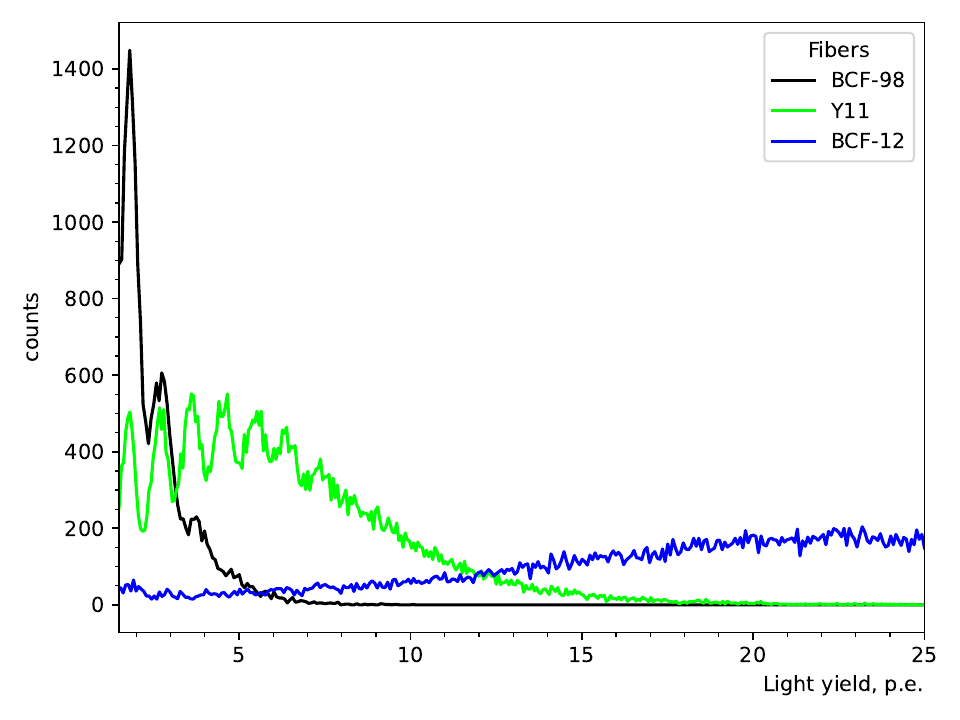}
\includegraphics[width=0.5\textwidth]{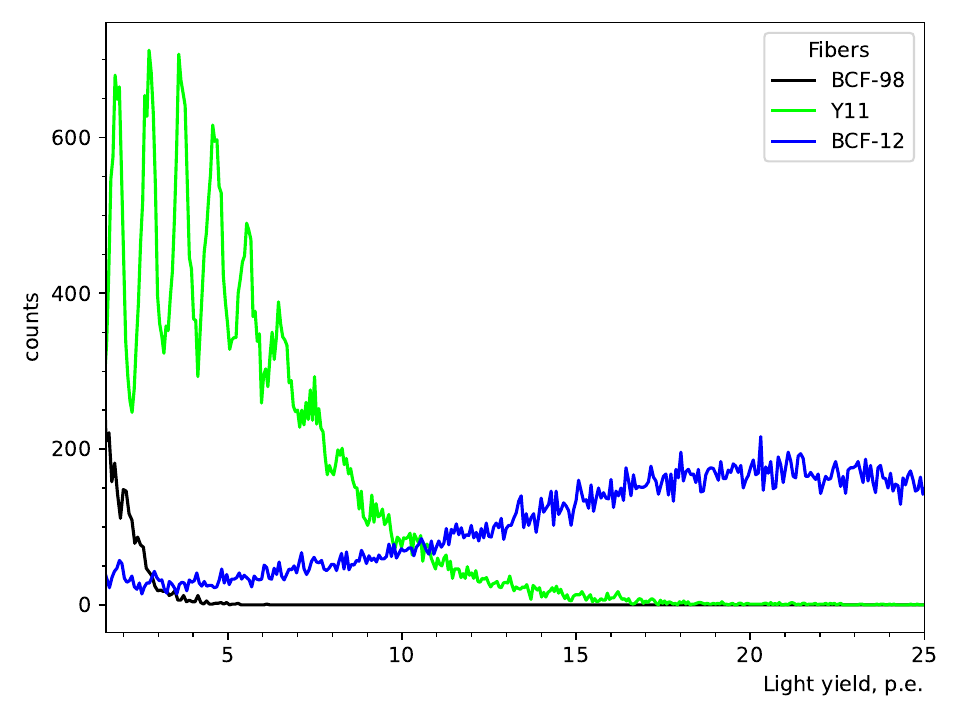}
\caption{\label{fig:45_and_135_degree}  Light yield in different fibers from a triggered $^{90}Sr$ radioactive source for 45$^{\circ}$(top) and  135$^{\circ}$ (bottom) crossing angles. The black, green, and blue histograms correspond to the clear, WLS, and scintillating fibers. The numbers of entries for the clear and WLS fibers in the range from $-$2~p.e. to 200~p.e. were normalized  to the  number of entries for the scintillating fiber and the pedestal peak observed for the scintillating fiber was subtracted. The distributions are shown above 1.5~p.e.}
\end{figure}

In order to detect the Cherenkov light, the measurements were performed with fibers at 45$^{\circ}$ degrees with respect to the average electron track direction in the collimator. The SiPM was either at 45$^{\circ}$  or  at 135$^{\circ}$ degree angles with respect to the track direction.
The obtained LY distributions are shown in Fig.~\ref{fig:45_and_135_degree}. 

 A sizable signal with an average value of $1.04\pm0.09$~p.e. is observed in the clear fiber at 45$^{\circ}$, while practically no signal ($0.09\pm0.06$~p.e.) is detected at 135$^{\circ}$.
This is clear evidence of Cherenkov light. 
Average LY for all measurements is summarized in Table~\ref{tab:all_angles}.
Statistical errors in this table are negligible. However, there are systematic uncertainties estimated from the multiple measurements of the same fiber, as discussed above. They are shown in the table.
Results for all 3 WLS fibers coincide within errors. The ratio of LY from the WLS and scintillating fibers, averaged over 3 angles and 3 WLS fiber samples, 
is 0.23$\pm$0.01.  
LY  also depends on the quality of fiber polishing. This source of systematic uncertainty was estimated by comparing the results obtained for the two ends of the same fiber. For the 3 WLS fibers, the ratios were 1.07, 0.99, and 1.17, respectively. For the scintillating fiber, the ratio was 0.89. The average of these 4 measurements was $1.03\pm0.07$. 
We conclude that the difference in the LY due to the differing quality of polishing is comparable to the reproducibility of the measurements and can be caused by imperfect reproducibility.
However, to be conservative, we included this additional  systematic uncertainty of 7\% to the final ratio of light yields from the WLS and scintillating fibers.
Adding in quadratures the  possible systematic uncertainties from the quality of the fiber polishing (7\%) and the systematic uncertainties estimated from the reproducibility of the measurements (6\%) we get the total systematic uncertainty  of 9\%. Therefore  the final result for the ratio of light yields from the WLS and scintillating fibers is $0.23\pm0.02$.
The light yield produced directly by charged particles in the WLS fibers is quite large, and it is desirable to include this effect in detector simulations.
The contribution of the Cherenkov light to the WLS and scintillating fiber signals is difficult to estimate. Cherenkov photons with short wavelength's can be absorbed by the dyes in the scintillating and WLS fibers and re-emitted in all directions. 
However, not all Cherenkov photons are captured by the Y11 dyes including photons produced in the cladding. Therefore the light yield at 45$^{\circ}$ is expected to be larger than the light yield at 135$^{\circ}$. The measurements agree with these expectations (see Table~\ref{tab:all_angles}) but the difference in the light yield is smaller than two standard deviations.
The increase of LY in the WLS and scintillating fibers at 135$^{\circ}$ in comparison with 90$^{\circ}$ is 1.24$\pm$0.1 and 1.16$\pm$0.18 correspondingly. The amplitude spectrum is expected to be shifted by a factor of 1.41 to the higher energies due to the increase of the track length in the fiber. However, the trigger threshold was the same at different angles. Therefore, the increase of the mean of the spectrum above the threshold is smaller than 1.41. Probably this is a reason for a slightly smaller LY increase at 135$^{\circ}$ in comparison with the expectation of 1.41. The same arguments are also valid for the case of 45$^{\circ}$, although the possible contribution from the Cherenkov light complicates the comparison.
The LY in the WLS and scintillating fibers is produced by two mechanisms: scintillation in the polystyrene with light absorption and re-emission by fiber dyes and the Cherenkov radiation which is also partially absorbed and re-emitted. The Cherenkov radiation contribution is hard to estimate without very complicated simulations. However, it can not be dominant even for WLS fibers. The LY in the clear fiber is more than 5 times smaller than the LY in the WLS fibers at the 45$^{\circ}$ crossing angle and negligible at 90$^{\circ}$ and 135$^{\circ}$. Moreover the Cherenkov light contribution to the organic scintillator signals is usually at a few per-cent level (see e.g.~\cite{LiquidO:2025qia}).
The BCF-12 fiber is a single-clad one while Y11 fibers are double-clad. Therefore the BCF-12 fiber has a smaller internal scintillation light trapping efficiency of 3.44\%~\cite{sheet1} in comparison with 5.4\% in case of the Y11 fibers~\cite{sheet2}. The SiPM light detection efficiencies are practically the same for the Y11 and BCF-12 fibers although they have quite different emission spectra (see Fig.~\ref{fig:spectra}). These efficiencies are 36.5\% and 38.8\% respectively. Quantitative comparison of light production in these fibers requires very complicated simulations. A rough estimate can be made assuming that the energy release for the given track length is the same in Y11 and BCF-12 fibers, since the material of the core is polystyrene in both cases. The attenuation lengths and the core radii are practically equal as well. The number of detected p.e. is proportional to the number of produced scintillation photons multiplied by the SiPM  efficiency, which depends on the fiber emission spectrum, and the fiber trapping efficiency. The contribution of non-absorbed Cherenkov photons is neglected in this estimate. 
The number of produced photons per MeV in the Y11 fibers
is a product of the observed ratio of LYs in the Y11 and BCF-12 fibers (see Table~\ref{tab:all_angles}), the inverse ratio of SiPM efficiencies and trapping efficiencies for these fibers, and  the number of produced photons per MeV  in the BCF-12 fibers which is about 8000~\cite{sheet1}.
Hence the number of photons per MeV in the Y11 fibers is about 1200 photons per MeV. This estimate inherits the 9\% error of the LY ratio, however the final uncertainty could be dominated by the uncertainty in the number of produced photons per MeV in the BCF-12 fibers which is not given by the manufacturer.

\begin{table}[htbp]
\centering
\caption{\label{tab:all_angles} Average LY in p.e. from different fiber types at three electron crossing angles.}
\smallskip
\begin{tabular}{|l|c|c|c|}
\hline
Fiber type & 90$^{\circ}$ & 45$^{\circ}$ & 135$^{\circ}$\\
\hline
BCF-12 &     $18.45\pm1.94$ &      $21.72\pm1.63$ &       $21.38\pm2.47$ \\
BCF-98 &      $0.11\pm0.01$ &       $1.04\pm0.09$ &        $0.09\pm0.06$ \\
Y11(200)MSJ&      $3.59\pm0.44$ &       $5.11\pm0.50$ &        $4.68\pm0.31$\\
Y11(200)MS&      $3.72\pm0.45$ &       $5.69\pm0.90$ &        $4.89\pm0.69$ \\
Y11(200)MS&      $4.03\pm0.38$ &       $5.65\pm0.42$ &        $4.78\pm0.62$ \\
Average Y11&    $3.81\pm0.24$ & $5.45\pm0.30$ & $4.73\pm0.26$\\
Y11/BCF-12&    $0.21\pm0.03$ & $0.25\pm0.02$ & $0.22\pm0.03$\\
\hline
\end{tabular}
\end{table}

\section{Conclusions}

Charged particles directly produce sizable light signals in the Kuraray Y11(200) 1~mm diameter WLS fibers. No difference was observed between Y11(200)MSJ and Y11(200)MS fibers. 
The measured LY in the 3 samples of the Y11 fibers expressed in the number of detected p.e. is as large as $23\pm2~\%$ of the LY of the Bicron BCF-12 scintillating fiber. Using the measured ratio of LY for two types of fibers a naive estimate gives about 1200 photons per MeV in the Y11 fiber. This estimate takes into account the known light production in BCF-12 fibers (8000 photons per MeV), the light trapping efficiencies of Y11 and BCF-12 fibers (5.4\% and 3.44\%, respectively) and the SiPM detection efficiencies (36.5\% and 38.8\%, respectively), but neglects the influence from non-absorbed Cherenkov photons. 
Practically no scintillation light was observed in the Bicron BCF-98 1~mm diameter clear fiber, but the Cherenkov light was clearly observed at the 45$^{\circ}$ track crossing angle. The contribution of the Cherenkov light to the signals observed in the Y11 WLS fibers is difficult to estimate since the directionality of the Cherenkov light is partially lost due to the absorption and re-emission of light by the dyes. 
It is desirable to include the direct light production by charged particles in the WLS fibers in detector Monte Carlo simulations, especially for detectors with good  energy or time resolution. For example the inclusion of direct signals from WLS fibers in simulations is expected to reduce the SFGD scintillating cubes~\cite{alekseev2026precisionlightyieldcrosstalk} response non-uniformity in the vicinity of the fiber by about 20\%. Similar improvements in the response uniformity are expected in the new scintillation counters of the DANSS experiment~\cite{DANSS:2021raa}.

\section*{Acknowledgments}
This work is supported in part by the Russian Science Foundation grant number 24-72-10089.


\bibliographystyle{elsarticle-num}






\end{document}